\documentclass{pnastwo}

\usepackage{graphicx}

\usepackage{hyperref}

\usepackage{amssymb,amsfonts,amsmath}

\copyrightyear{2015}

\begin{document}


\title{Continuous voting by approval and participation} 






\author{Renato Fabbri\affil{2}{University of S\~ao Paulo}
\and
Ricardo Poppi\affil{1}{Brazilian Presidency of the Republic}
}

\contributor{Draft}


\maketitle 

\begin{article}


\begin{abstract}
In finding the adequate way to prioritize proposals, the Brazilian participation community agreed about the measurement of two indexes, one of approval and one of participation.
Both practice and literature is constantly handled by the experts involved, and the formalization of such model and metrics seems novel.
Also, the relevance of this report is strengthened by the nearby use of these indexes by the Brazilian General Secretariat of the Republic to raise and prioritize proposals about public health care in open processes.
\end{abstract}


\keywords{social participation | recommendation systems | online voting | statistics} 






\section{}
Online decision making is a kind of recommendation system with special appeal for online social participation and electronic governments. This poses challenges on the design of such
processes regarding validity, security and the adequate indicators. Indeed,
the processes themselves vary, and the fact that the indexes presented here seem not to be
formalized and published is an evidence that such online decision making is very recent phenomena.

The main contribution of this report is a modeling for
an online voting process~\cite{issue1,issue2,tabRonald} with the following characteristics:
\begin{itemize}
    \item proposals might be inserted by voters after the voting phase started.
    \item Voting might be extended as a permanent process. In other words, voting on and adding new proposals might be open continuously.
    \item A proposal is presented to a voter one by one as random outcomes of all proposals.
    \item Each vote might be of one and only type among: ``approve'', ``disapprove'' and ``indifferent''.
    \item Voters vote without authentication.
    \item Intended mostly for national rankings, but can also be local or have foreign participation.
    \item Should result in a ranking of proposals to assist public management.
\end{itemize}

This setting requires care about security and validity.
Some of which are:
\begin{itemize}
	\item adequate sampling of individual proposals and overall ranking.
    \item Registration of the IP address and time of votes to ease detection of automated and other fraudulent efforts.
    \item Reasonable use of the outcomes from the voting process. This requires probing the survey being conducted and its purposes. The indexes here presented target indicatives for the Brazilian federal government about the most important health care proposals. Given the unauthenticated voting, the outcomes might be regarded as reference rankings if data is minimally shared and checked for inadequate data entry (such as voting by automated scripts or a persistent participant introduced bias).
\end{itemize}


\section{}

\subsection{Approval and participation indexes}

The approval index $\alpha_i$ and the participation index $\gamma_i$ of the proposal $i$ was defined as:
\begin{equation}
\begin{split}
\alpha_i=\frac{v^+_i-v^-_i}{\eta_i} \\
\gamma_i=\frac{v^+_i+v^-_i}{\eta_i}
\end{split}
\end{equation}
where $v^+_i$, $v^-_i$ and $\eta_i$ are approval count, disapproval count and exhibition count, respectively.
Note that $\alpha_i\in[-1,1]$, $\gamma_i\in[0,1]$, and $v^o_i=\eta_i-v^+_i-v^-_i$ is the count of the ``indifferent'' manifestations received by proposal $i$.
Also, such $\alpha_i$ and $\gamma_i$ indexes are expected, for each proposal $i$, to be a constant  plus a sampling estimate error that should be smaller as $\eta_i$ raises. This error is thought to be acceptable if $\eta_i$ is above a threshold $\overline{\eta}$ established by the participation community and public managers.
As an initial decision, the staff agreed to use $\overline{\eta}$ as to select $10-20\%$ of all proposals.
A threshold $\overline{\gamma}$ can be used as a required level of engagement for proposals to be relevant,
while the threshold $\overline{\alpha}$ is used to classify the outcome
as ``approved'',``disapproved'' and ``clash''. More specifically:

\begin{equation}
\begin{split}
\eta_i > \overline{\eta}    \;\;\;\;\;\; & \Rightarrow\;\;\;\;\;\; i     \text{ is sampled} \\
\gamma_i>\overline{\gamma}    \;\;\;\;\;\; & \Rightarrow\;\;\;\;\;\; i    \text{ is relevant} \\
|\alpha_i| \leq\overline{\alpha}\;\;\;\;\;\; & \Rightarrow\;\;\;\;\;\; i  \text{ is a clash} \\
\alpha_i > \overline{\alpha}  \;\;\;\;\;\; & \Rightarrow\;\;\;\;\;\; i    \text{ is approved} \\
-\alpha_i > \overline{\alpha} \;\;\;\;\;\; & \Rightarrow\;\;\;\;\;\; i   \text{ is disapproved} \\
\end{split}
\end{equation}

If a proposal is both sampled and relevant, than it is prioritized.
The coherent values of $\overline{\alpha_i}=0.5$ (or $1/3$)
and $\overline{\gamma_i}=0.5$ were chosen as standards of the decision model. 
These are likely to change with implementation and management.

Thresholds might be dependent on proposal, such as given by meaningful expressions.
Immediate examples are:

\begin{equation*}
\begin{split}
\overline{\gamma_i} & =1-\frac{\eta_i}{max(\{\eta_j , \;\forall\; j \})} \\
\overline{\alpha_i} & = 1-\gamma_i
\end{split}
\end{equation*}
\noindent These bonds among proposal variables has been discarded by the staff in the present initial steps.


\vfill
\vspace{.6cm}
\vfill

\subsection{Selected decision framework examples}

Many of the online decision processes conceived and practiced resemble our model and have similar measurements to the $\alpha_i$ and $\gamma_i$ indexes. This section presents a collection of models more familiar to the Brazilian participatory community, with focus on the mechanisms, not on historical notes.

Pairwise~\cite{pairwise} is part of the tackled paradigm: the ranking procedure accepts new proposals while the voting occurs. Even so, pairwise voting is comparative, voter chooses between two proposals at each vote, and this does not fit proposed procedure.

Appgree software~\cite{appgree} ranks proposals by sampling voters in cycles, each with fewer proposals. This is
adequate for a range of decision making cases and showcases statistical estimates utility. The system has a separate proposition phase, and relies on an organized group engagement and user identities, which also does not fit current needs. 

Liquid Feedback~\cite{liquid} is a very renowned and bleeding edge solution for collective decision making.
It relies on delegating your voting count on specific subjects to other people you know or trust.
Therefore, it does not fit current needs. Even so, this framework have precious considerations for our case, such as about ranking and presenting proposals to voters in the most useful ways.

A Brazilian solution, used in diverse software and specially important as the output of a nation-wide decision making need, is the Agora Algorithm~\cite{agora}. It presents a decision procedure in phases (agenda proposition, deliberations proposition and commenting, voting) with resolution outcomes.
Although coherent, this framework requires authentication and might need experimentation and tuning in order to be effective with more than dozens or a few hundreds of participants.

There is a number of other solutions for online collaborative prioritization, such as IdeaScale, Kidling, or any flavor of an Analytic Hierarchy Process (AHP). Authors hope to better formalize possible solutions (and found implementations), maybe through recommender systems theory~\cite{coursera}.


%

\section{Discussion}
The above estimates are the best fit the researchers could deliver,
suitable for current needs and not found (yet) in literature.
The following questions should be addressed in near future:
\begin{itemize}
	\item Are there more adequate metrics for ranking proposals in the given setting?
    \item What are strong and weak aspects of the approach for collective recommendation?
    \item What thresholds will be the choice of community and will they be adjusted with time?
    \item Are there really no previous formalized model of this setting? If there is, what comparisons can we make on design, metrics and outcomes?
    \item To which extent will participation community and public managers legitimize this approach?
	\item What is the impact of this technological approach in public health care, social participation and the scientific community?
    \item To which extent society benefits from this continuous voting process? Is it worth the time spent by voters? How to evaluate this relation in terms of spent and gained resources?
\end{itemize}

Most importantly, this report is being delivered to the civil society and the scientific community for consideration. Given the large number of possibilities for the collective ranking procedure, and the proliferation of solutions, research efforts might aim the organization of such procedures.




%


\begin{acknowledgments}
	Author is grateful to CNPq (process
	140860/2013-4, project 870336/1997-5), UNDP
	(contract 2013/00056, project BRA/12/018), SNAS/SGPR,
	and the Postgraduate Committee of the IFSC/USP.
	Thanks to the Brazilian social participation community for the conception and practice of this specific voting setting.
\end{acknowledgments}







\end{article}

\end{document}